\documentclass[12pt,onecolumn]{IEEEtran}

\ifCLASSINFOpdf
   \usepackage[pdftex]{graphicx}

\else

\fi

\usepackage{color}
\usepackage[cmex10]{amsmath}
\usepackage{amssymb}   
\usepackage{amsxtra}
\usepackage{amscd}
\usepackage{amsthm}

\usepackage{graphicx}   

\usepackage{array}  

\usepackage{multirow}  

 \usepackage{flushend}

\usepackage{cite}

\usepackage{color,soul} 
\soulregister\cite7
\soulregister\ref7
\soulregister\pageref7

\begin{document}
%
\title{{\huge Index Modulation Techniques for 5G Wireless Networks}}

%
%
%

\author{Ertugrul~Basar,~\IEEEmembership{Senior Member,~IEEE} 
	\thanks{\textit{E. Basar is with Istanbul Technical University, Faculty of Electrical and Electronics Engineering, 34469, Istanbul, Turkey.} }
	\thanks{\textit{e-mail: basarer@itu.edu.tr}}

	\thanks{\textit{This work is supported in part by the Scientific and Technological Research Council of Turkey (TUBITAK) under Grant no. 114E607.}}
		\thanks {\textit{Copyright (c) 2016 IEEE. Personal use of this material is permitted. However, permission to use this material for any other purposes must be obtained from the IEEE by sending a request to pubs-permissions@ieee.org.}}
	
	}

%
%

\markboth{IEEE Communications Magazine}{Ba\c{s}ar:  Spatial and Index Modulation Techniques for 5G Wireless Networks}

%




\maketitle

\begin{abstract}

 The increasing demand for higher data rates, better quality of service, fully mobile and connected wireless networks lead the researchers to seek new solutions beyond 4G wireless systems. It is anticipated that 5G wireless networks, which are expected to be introduced around 2020, will achieve ten times higher spectral and energy efficiency than current 4G wireless networks and will support data rates up to $10$ Gbps for low mobility users. The ambitious goals set for 5G wireless networks require dramatic changes in the design of different layers for next generation communications systems. Massive multiple-input multiple-output (MIMO) systems, filter bank multi-carrier (FBMC) modulation, relaying technologies, and millimeter-wave communications have been considered as some of the strong candidates for the physical layer design of 5G networks. In this article, we shed light on the potential and implementation of index modulation (IM) techniques for MIMO and multi-carrier communications systems which are expected to be two of the key technologies for 5G systems. Specifically, we focus on two promising applications of IM: spatial modulation (SM) and orthogonal frequency division multiplexing with IM (OFDM-IM), and we discuss the recent advances and future research directions in IM technologies towards spectral and energy-efficient 5G wireless networks.

\end{abstract}
\vspace*{-0.1cm}

%
\IEEEpeerreviewmaketitle

\renewcommand{\thefootnote}{\arabic{footnote}}

\section{Introduction}
After more than 20 years of research and development, the achievable data rates of today's cellular wireless communications systems are several thousands of times faster compared to earlier 2G wireless systems. However, unprecedented levels of spectral and energy efficiency are expected from 5G wireless networks to achieve ubiquitous communications between anybody, anything, and anytime \cite{5G2}. In order to reach the challenging objectives of 5G wireless networks, the researchers have envisioned novel physical layer (PHY) concepts such as massive multiple-input multiple-output (MIMO) systems and non-orthogonal multi-carrier communications schemes. However, the wireless community is still working day and night to come up with new and more effective PHY solutions towards 5G networks. There has been a growing interest on index modulation (IM) techniques over the past few years. IM, in which the indices of the building blocks of the considered communications systems are used to convey additional information bits, is a novel digital modulation scheme with high spectral and energy efficiency. \textit{Spatial modulation (SM)} and \textit{orthogonal frequency division multiplexing with IM (OFDM-IM)} schemes, where the corresponding index modulated building blocks respectively are the transmit antennas of a MIMO system and the subcarriers of an OFDM system, appear as two interesting as well as promising applications of the IM concept. 

After the pioneering works of Mesleh \textit{et al.} \cite{SM_jour} and Jeganathan \textit{et al.} \cite{SSK}, SM techniques have attracted significant attention over the past few years. Although having strong and well-established competitors such as vertical Bell Labs layered space-time (V-BLAST) and space-time coding (STC) systems, SM schemes have been regarded as possible candidates for spectral and energy-efficient next generation MIMO systems. On the other hand, the researchers have started to explore the potential of IM concept for the subcarriers of OFDM systems in recent times and it has been shown that the OFDM-IM scheme \cite{OFDM_IM} can provide attractive advantages over classical OFDM, which is an integral part of many current wireless standards. 

The aim of this article is to present the basic principles of these two promising schemes, SM and OFDM-IM, which are still waiting to be explored by many experts, and review some of the recent interesting results in IM techniques. Furthermore, we discuss the implementation scenarios of IM techniques for next generation wireless networks and outline possible future research directions. Particularly, we shift our focus to generalized, enhanced, and quadrature IM schemes and the application of IM techniques for massive multi-user MIMO (MU-MIMO) and cooperative communications systems.

\section{Index Modulation for Transmit Antennas: Spatial Modulation}
SM is a novel way of transmitting information by means of the indices of the transmit antennas of a MIMO system in addition to the conventional $M$-ary signal constellations. In contrast to conventional MIMO schemes which rely either on spatial multiplexing to boost the data rate or spatial diversity to improve the error performance, the multiple transmit antennas of a MIMO system are used for a different purpose in an SM scheme. More specifically, there are two information carrying units in SM: indices of transmit antennas and $M$-ary constellation symbols. For each signaling interval, a total of
\begin{equation}
\log_2(n_T) + \log_2(M)
\end{equation}
bits enter the transmitter of an SM system as seen from Fig. 1, where $n_T$ and $n_R$ denote the number of transmit and receive antennas, respectively, and $M$ is the size of the considered signal constellation such as $M$-ary phase shift keying ($M$-PSK) or $M$-ary quadrature amplitude modulation ($M$-QAM). The $\log_2(M)$ bits of the incoming bit sequence are used to modulate the phase and/or amplitude of a carrier signal traditionally, while the remaining $\log_2(n_T)$ bits of the incoming bit sequence are reserved for the selection of the index $(I)$ of the active transmit antenna which performs the transmission of the corresponding modulated signal $(s)$.

The receiver of the SM scheme has two major tasks to accomplish: detection of the active transmit antenna for the demodulation of the index selecting bits and detection of the data symbol transmitted over the activated transmit antenna for the demodulation of the bits mapped to $M$-ary signal constellation. Unfortunately, the optimum maximum likelihood (ML) detector of SM has to make a joint search over all transmit antennas and constellation symbols to perform these two tasks. In other words, the ML detector of the SM scheme independently implements a classical single-input multiple-output (SIMO) ML detector for all transmit antennas to find the activated transmit antenna by comparing the minimum decision metrics $(m_1,m_2,\ldots,m_{n_T})$. On the other hand, the primitive suboptimal detector of SM deals with the aforementioned two tasks one by one, that is, first, it determines the activated transmit antenna, second, it finds the data symbol transmitted over this antenna. Therefore, the size of the search space becomes $n_T \times M$  and $n_T + M$ for the ML and suboptimal detector, respectively.

SM systems provide important advantages over classical MIMO systems, which are extensively covered in the literature\cite{SM_magazine_2,Design_SM}. The main advantages of SM over classical MIMO systems can be summarized as follows:   
\begin{itemize}
\item Simple transceiver design: Since only a single transmit antenna is activated, a single radio frequency (RF) chain can handle the transmission for the SM scheme. Meanwhile, inter-antenna synchronization (IAS) and inter-channel interference (ICI) are completely eliminated, and the decoding complexity of the receiver, in terms of total number of real multiplications performed, grows linearly with the constellation size and number of transmit antennas.
\item Operation with flexible MIMO systems: SM does not restrict the number of receive antennas as the V-BLAST scheme, which requires $n_R>n_T$ to operate with minimum mean square error (MMSE) detector.
\item High spectral efficiency: Due to the use of antenna indices as an additional source of information, the spectral efficiency of SM is higher than that of single-input single-output (SISO) and orthogonal STC systems. 
\item High energy efficiency: The power consumed by the SM transmitter is independent from number of transmit antennas while information can be still transfered via these antennas. Therefore, SM appears as a green and energy-efficient MIMO technology.
 
\end{itemize}  

As an example, assuming an $n_T \times n_R$ MIMO system operating  at a fixed spectral efficiency, SM achieves $200(n_T-1)/(2n_T+1)\%$ reduction in ML detection complexity (in terms of total number of real multiplications) compared to V-BLAST due to the activation of a single transmit antenna. Furthermore, the sparse structure of the transmission vectors allows the implementation of several near/sub-optimal low-complexity detection methods for SM systems such as matched filter based detection and compressed sensing based detection. In terms of the energy efficiency in Mbits/J, improvements up to $46\%$ compared to V-BLAST are reported for different type of base stations (BSs) equipped with multiple antennas.

While the SM scheme has the aforementioned appealing advantages, it also has some disadvantages. The spectral efficiency of SM increases logarithmically with $n_T$, while the spectral efficiency of V-BLAST increases linearly with $n_T$. Therefore, higher number of transmit antennas are required for SM to reach the same spectral efficiency as that of V-BLAST. The channel coefficients of different transmit antennas must be sufficiently different for an SM scheme to operate effectively. In other words, SM requires rich scattering environments to ensure better error performance.  Since SM transfers the information using only the spatial domain, plain SM cannot provide transmit diversity as STC systems which rely on both spatial and time domains for data transmission.

Considering the advantages and disadvantages of SM systems, we may conclude that SM provides an interesting trade-off among complexity, spectral efficiency, and error performance. Consequently, SM has been regarded as a possible candidate for spectral and energy-efficient next generation wireless communications systems \cite{5G2}.

\begin{figure}[t]
	\begin{center}
		{\includegraphics[scale=0.8]{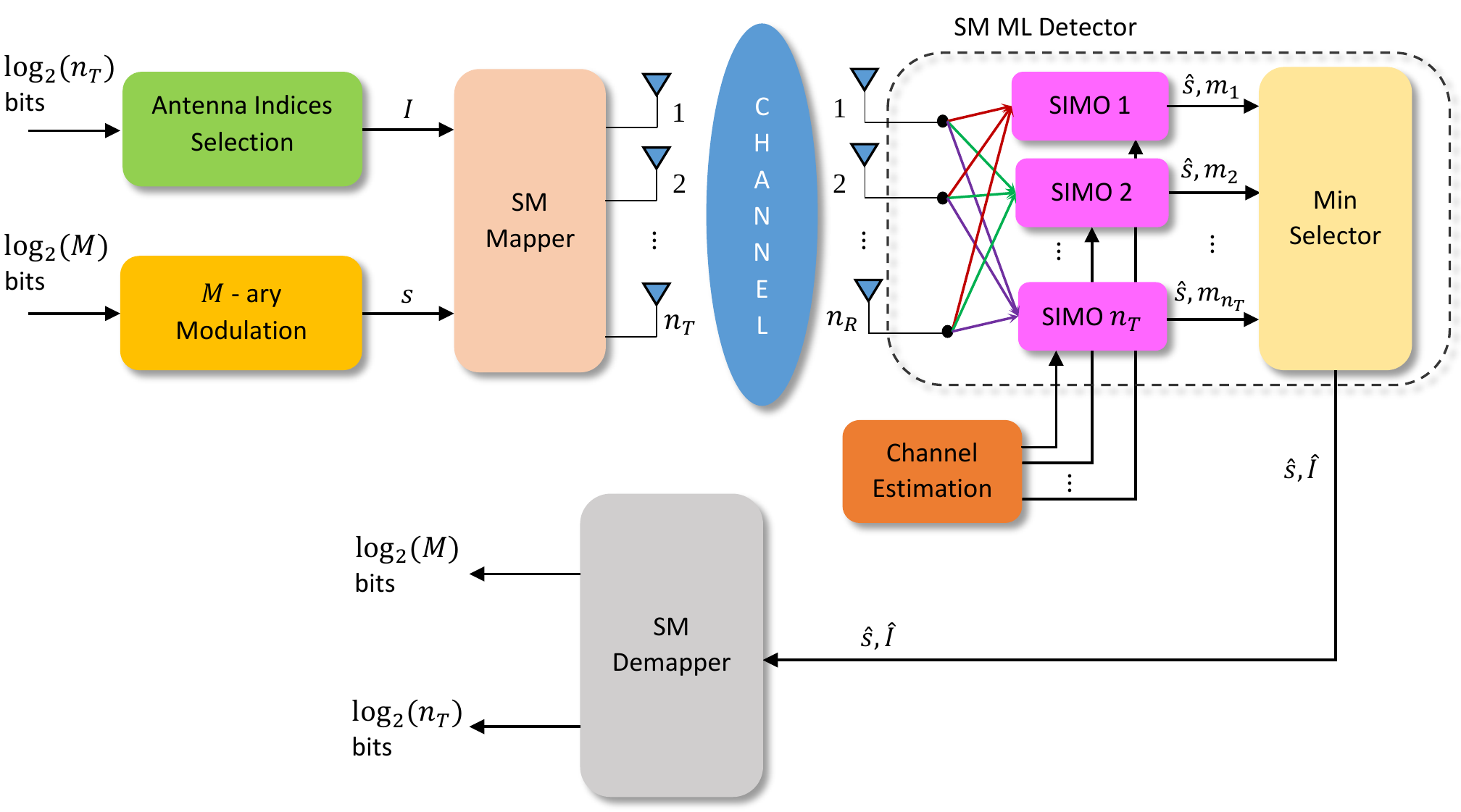}}
		\vspace*{-0.2cm}
		\caption{Block diagram of the SM transceiver for an $n_T \times n_R$ MIMO system. $s\,(\text{or }\hat{s})$ and $I\,(\text{or }\hat{I}) \in \left\lbrace 1,2,\ldots,n_T \right\rbrace $ denote the selected (or estimated) $M$-ary constellation symbol and transmit antenna index, respectively and $m_n,n=1,2,\ldots,n_T$ is the minimum metric provided by the $n$th SIMO ML detector.}
		\vspace*{-0.4cm}
	\end{center}
\end{figure}

\section{Recent Advances in SM}

The first studies on SM concept date back to the beginning of 2000s in which the researchers used different terminologies. However, after the inspiring works of Mesleh \textit{et al.} \cite{SM_jour} and Jeganathan \textit{et al.} \cite{SSK}, numerous papers on SM have been published in which the experts focus on generalized, spectral and energy-efficient SM systems, low-complexity detector types, block/trellis coded SM systems with transmit/time diversity, link adaptation methods such as adaptive modulation, transmit antenna selection and precoding, performance analysis for different fading channel types and channel estimation errors, information theoretical analyses, differential SM schemes with non-coherent detection, cooperative SM systems, and so on. For a comprehensive overview of these studies, interested readers are referred to survey papers \cite{SM_magazine_2} and \cite{Design_SM}.  

In this section, we review some of the recent as well as promising advances in SM technologies such as generalized, enhanced, and quadrature SM systems, massive MU-MIMO systems with SM, and cooperative SM schemes, which have the potential to provide efficient solutions towards 5G wireless networks.

\subsection{Generalized, Enhanced, and Quadrature SM Schemes}
As mentioned earlier, the major disadvantage of SM is its lower spectral efficiency compared to classical V-BLAST scheme for the same number of transmit antennas. Although a considerable number of information bits can still be transmitted by the indices of active transmit antennas, for higher order modulations and MIMO systems, SM suffers a significant loss in spectral efficiency with respect to V-BLAST due to its inactive transmit antennas.    

One of the first attempts to not only increase the spectral efficiency of SM but also ease the constraint on number of transmit antennas, which has to be an integer power of two for classical SM, has been made in \cite{GSM2} by the generalized SM (GSM) scheme, where the number of active transmit antennas is no longer fixed to unity. In the GSM scheme, multiple active transmit antennas are selected to transmit the same data symbol. Denoting the number of active transmit antennas by $n_A$ where $n_A<n_T$, $ \lfloor \log_2 \binom{n_T}{n_A} \rfloor  $ information bits can be transmitted for each signaling interval in addition to the $\log_2(M)$ bits transmitted by the $M$-ary data symbols, where $\lfloor \cdot \rfloor$ is the floor operation. Since $ \log_2(n_T) \le \lfloor \log_2 \binom{n_T}{n_A} \rfloor $ for $n_T=2^n \, (n=1,2,\ldots)$, the spatial domain can be used more effectively by the GSM scheme. As an example, for $n_T=8$, only three bits can be transmitted by the antenna indices in SM, while this can be doubled by GSM for $n_A=4$. In \cite{GSM_MA_SM}, the concept of GSM has been extended to multiple-active spatial modulation (MA-SM), where different data symbols are transmitted from the selected transmit antennas to further boost the spectral efficiency. Therefore, the spectral efficiency of the MA-SM scheme becomes $ \lfloor \log_2 \binom{n_T}{n_A} \rfloor  + n_A \log_2(M)  $ bits per channel use (bpcu), which is considerably higher than that of SM. It should be noted that MA-SM provides an intermediate solution between two extreme schemes: SM and V-BLAST, which are special cases of MA-SM with $n_A=1$ and $n_A=n_T$, respectively.

Enhanced SM (ESM) is a recently proposed and promising variant of SM \cite{ESM}. In the ESM scheme, the number of active transmit antennas can vary for each signaling interval and the information is conveyed not only by the indices of active transmit antennas but also by the selected signal constellations used in transmission. In other words, the ESM scheme considers multiple signal constellations and the information is transmitted by the combination of active transmit antennas and signal constellations. As an example, for two transmit antennas and four bpcu transmission, the ESM scheme transmits two bits by the joint selection of active transmit antennas and signal constellations, where one quadrature PSK (QPSK) and two binary PSK (BPSK) signal constellations (one ordinary and one rotated) can be  used \cite{ESM}. For two-bit sequences $\left\lbrace 0,0 \right\rbrace $, $\left\lbrace 0,1 \right\rbrace $, $\left\lbrace 1,0 \right\rbrace $, and $\left\lbrace 1,1 \right\rbrace $, the ESM scheme uses the following transmission vectors, respectively: $ \begin{bmatrix}
\mathcal{S}_4 & 0
\end{bmatrix}^\textrm{T} $, $ \begin{bmatrix}
0 & \mathcal{S}_4 
\end{bmatrix}^\textrm{T} $, $ \begin{bmatrix}
\mathcal{S}_2 & \mathcal{S}_2
\end{bmatrix}^\textrm{T} $, and $ \begin{bmatrix}
\mathcal{S}_2 e^{j\theta} & \mathcal{S}_2 e^{j\theta}
\end{bmatrix}^\textrm{T} $, where $\mathcal{S}_m,m=2,4$ denotes $M$-PSK constellation, $(\cdot)^\textrm{T}$ stand for transposition of a vector and $\theta=\pi/2$ is a rotation angle used to obtain a third signal constellation in addition to classical BPSK and QPSK signal constellations. Other implementation examples of ESM can be found in \cite{ESM}.  It is interesting to that the first two transmission vectors of the ESM scheme correspond to the classical SM using QPSK with single activated transmit antenna, where the first and second transmit antenna is used for the transmission of a QPSK symbol, respectively. On the other hand, the third and fourth transmission vectors correspond to the simultaneous transmission of two symbols selected from BPSK and modified BPSK constellations, respectively. The reason behind reducing the constellation size from four to two can be explained by the fact that same number of information bits (two bits for this case) must be carried with $M$-ary constellations independent from number of active transmit antennas. Examples of the generalization of the ESM scheme for different number of transmit antennas and signal constellations can be found in \cite{ESM}.   

Quadrature SM (QSM) \cite{QSM} is yet another clever modification of classical SM to improve the spectral efficiency while maintaining its advantages such as operation with single RF chain and ICI free transmission. In the QSM scheme, the real and imaginary parts of the complex $M$-ary data symbols are separately transmitted using the SM principle. For a MIMO system with $n_T$ transmit antennas, the spectral efficiency of QSM becomes $2\log_2(n_T) + \log_2(M)$ bpcu by simultaneously applying SM principle for in-phase and quadrature components of the complex data symbols. 
As an example, for $n_T=2$ and $M=4$, in addition to the two bits mapped to the QPSK constellation, extra two bits can be transmitted in the spatial domain by using one of the following four transmission vectors: $ \begin{bmatrix}
s_R+js_I & 0
\end{bmatrix}^\textrm{T} $, $ \begin{bmatrix}
s_R & js_I 
\end{bmatrix}^\textrm{T} $, $ \begin{bmatrix}
js_I & s_R
\end{bmatrix}^\textrm{T} $, and $ \begin{bmatrix}
0 & s_R+js_I
\end{bmatrix}^\textrm{T} $ for input bit sequences $\left\lbrace 0,0 \right\rbrace $, $\left\lbrace 0,1 \right\rbrace $, $\left\lbrace 1,0 \right\rbrace $, and $\left\lbrace 1,1 \right\rbrace $, respectively, where $s_R$ and $s_I$ denote the real and imaginary parts of $s=s_R+js_I \in \mathcal{S}_4$ , respectively. It is interesting to note that the first and second element of these two-bit sequences indicates the position of the real and imaginary part of $s$, respectively. 
Even if the number of active transmit antennas can be one or two for the QSM scheme, a single RF chain is sufficient at the transmitter since only two carriers (cosine and sine) are used during transmission.

 \begin{table}[t]
 	
 	\begin{center}
 		\caption{Transmission vectors $\left( \mathbf{x}^T\right) $ of SM, ESM, and QSM schemes for $4$ bpcu and two transmit antennas $(n_T=2)$, red bits indicate the single bit transmitted by the spatial domain for SM, blue bits indicate the additional one bit transmitted by the spatial domain for ESM and QSM}
 		\begin{tabular}{|c|c|c|c||c|c|c|c|}
 			\hline \rule[-2ex]{0pt}{5.5ex} Bits & SM  & ESM  & QSM & Bits & SM & ESM & QSM  \\  \hline 
 			\hline \rule[-2ex]{0pt}{5.5ex} \textcolor{red}{0}\textcolor{blue}{0}00 & $\begin{bmatrix} 1 & 0 \end{bmatrix}$ & $\begin{bmatrix}  \frac{1+j}{\sqrt{2}} & 0  \end{bmatrix}$   & $\begin{bmatrix}  \frac{1+j}{\sqrt{2}} & 0  \end{bmatrix}$  &                    
 			\textcolor{red}{1}\textcolor{blue}{0}00 & $\begin{bmatrix} 0 & 1 \end{bmatrix}$   & $\begin{bmatrix}  \frac{1}{\sqrt{2}} & \frac{1}{\sqrt{2}}  \end{bmatrix}$ &  $\begin{bmatrix}  \frac{j}{\sqrt{2}} & \frac{1}{\sqrt{2}}  \end{bmatrix}$ \\ 
 			
 			\hline \rule[-2ex]{0pt}{5.5ex} \textcolor{red}{0}\textcolor{blue}{0}01 & $\begin{bmatrix} \frac{1+j}{\sqrt{2}} & 0  \end{bmatrix}$ & $\begin{bmatrix}  \frac{-1+j}{\sqrt{2}} & 0  \end{bmatrix}$   & $\begin{bmatrix}  \frac{-1+j}{\sqrt{2}} & 0  \end{bmatrix}$  & 
 			\textcolor{red}{1}\textcolor{blue}{0}01 & $\begin{bmatrix} 0& \frac{1+j}{\sqrt{2}}   \end{bmatrix}$ & $\begin{bmatrix}  \frac{1}{\sqrt{2}} & \frac{-1}{\sqrt{2}}  \end{bmatrix}$  &  $\begin{bmatrix}  \frac{j}{\sqrt{2}} & \frac{-1}{\sqrt{2}}  \end{bmatrix}$   \\ 
 			
 			\hline \rule[-2ex]{0pt}{5.5ex} \textcolor{red}{0}\textcolor{blue}{0}10 & $\begin{bmatrix} j & 0  \end{bmatrix}$ & $\begin{bmatrix}  \frac{-1-j}{\sqrt{2}} & 0  \end{bmatrix}$   & $\begin{bmatrix}  \frac{-1-j}{\sqrt{2}} & 0  \end{bmatrix}$  & 
 			\textcolor{red}{1}\textcolor{blue}{0}10 & $\begin{bmatrix} 0& j  \end{bmatrix}$ & $\begin{bmatrix}  \frac{-1}{\sqrt{2}} & \frac{1}{\sqrt{2}}  \end{bmatrix}$  &  $\begin{bmatrix}  \frac{-j}{\sqrt{2}} & \frac{-1}{\sqrt{2}}  \end{bmatrix}$   \\

 			\hline \rule[-2ex]{0pt}{5.5ex} \textcolor{red}{0}\textcolor{blue}{0}11 & $\begin{bmatrix} \frac{-1+j}{\sqrt{2}} & 0  \end{bmatrix}$ & $\begin{bmatrix}  \frac{1-j}{\sqrt{2}} & 0  \end{bmatrix}$   & $\begin{bmatrix}  \frac{1-j}{\sqrt{2}} & 0  \end{bmatrix}$  & 
 			\textcolor{red}{1}\textcolor{blue}{0}11 & $\begin{bmatrix} 0& \frac{-1+j}{\sqrt{2}}   \end{bmatrix}$ & $\begin{bmatrix}  \frac{-1}{\sqrt{2}} & \frac{-1}{\sqrt{2}}  \end{bmatrix}$  &  $\begin{bmatrix}  \frac{-j}{\sqrt{2}} & \frac{1}{\sqrt{2}}  \end{bmatrix}$   \\

 			\hline \rule[-2ex]{0pt}{5.5ex} \textcolor{red}{0}\textcolor{blue}{1}00 & $\begin{bmatrix} -1 & 0 \end{bmatrix}$ & $\begin{bmatrix} 0 & \frac{1+j}{\sqrt{2}}   \end{bmatrix}$   & $\begin{bmatrix}  \frac{1}{\sqrt{2}} & \frac{j}{\sqrt{2}}   \end{bmatrix}$  &                  
 			\textcolor{red}{1}\textcolor{blue}{1}00& $\begin{bmatrix} 0 & -1 \end{bmatrix}$ & $\begin{bmatrix}  \frac{j}{\sqrt{2}} & \frac{j}{\sqrt{2}}   \end{bmatrix}$ & $\begin{bmatrix} 0 & \frac{1+j}{\sqrt{2}}   \end{bmatrix}$   \\ 
 			
 			\hline \rule[-2ex]{0pt}{5.5ex} \textcolor{red}{0}\textcolor{blue}{1}01 & $\begin{bmatrix} \frac{-1-j}{\sqrt{2}} & 0  \end{bmatrix}$ & $\begin{bmatrix} 0 & \frac{-1+j}{\sqrt{2}}   \end{bmatrix}$   &  $\begin{bmatrix}  \frac{-1}{\sqrt{2}} & \frac{j}{\sqrt{2}}   \end{bmatrix}$ & 
 			\textcolor{red}{1}\textcolor{blue}{1}01& $\begin{bmatrix} 0 & \frac{-1-j}{\sqrt{2}}   \end{bmatrix}$ & $\begin{bmatrix}  \frac{j}{\sqrt{2}} & \frac{-j}{\sqrt{2}}   \end{bmatrix}$ &  $\begin{bmatrix} 0 & \frac{-1+j}{\sqrt{2}}   \end{bmatrix}$  \\ 
 			
 			\hline \rule[-2ex]{0pt}{5.5ex} \textcolor{red}{0}\textcolor{blue}{1}10 & $\begin{bmatrix} -j & 0 \end{bmatrix}$ & $\begin{bmatrix} 0 & \frac{-1-j}{\sqrt{2}}   \end{bmatrix}$ & $\begin{bmatrix}  \frac{-1}{\sqrt{2}} & \frac{-j}{\sqrt{2}}   \end{bmatrix}$  &                  
 			\textcolor{red}{1}\textcolor{blue}{1}10 & $\begin{bmatrix} 0 & -j \end{bmatrix}$ & $\begin{bmatrix}  \frac{-j}{\sqrt{2}} & \frac{j}{\sqrt{2}}   \end{bmatrix}$ &  $\begin{bmatrix} 0 & \frac{-1-j}{\sqrt{2}}   \end{bmatrix}$  \\ 
 			
 			\hline \rule[-2ex]{0pt}{5.5ex} \textcolor{red}{0}\textcolor{blue}{1}11 & $\begin{bmatrix} \frac{1-j}{\sqrt{2}} & 0 \end{bmatrix}$ & $\begin{bmatrix} 0 & \frac{1-j}{\sqrt{2}}   \end{bmatrix}$  &  $\begin{bmatrix}  \frac{1}{\sqrt{2}} & \frac{-j}{\sqrt{2}}   \end{bmatrix}$ & 
 			\textcolor{red}{1}\textcolor{blue}{1}11 & $\begin{bmatrix} 0 & \frac{1-j}{\sqrt{2}}  \end{bmatrix}$ &  $\begin{bmatrix}  \frac{-j}{\sqrt{2}} & \frac{-j}{\sqrt{2}}   \end{bmatrix}$ & $\begin{bmatrix} 0 & \frac{1-j}{\sqrt{2}}  \end{bmatrix}$  \\ 
 			\hline

 		\end{tabular} 
 	\end{center}
 	\vspace*{-0.3cm}
 \end{table}

In Table I, transmission vectors of SM, ESM, and QSM schemes are given for $4$ bpcu transmission and two transmit antennas, where we considered natural bit mapping for ease of presentation. We observe from Table I that both ESM and QSM schemes convey more bits by the spatial domain compared to conventional SM, which leads to not only improved spectral efficiency but also to higher energy efficiency.

In Fig. 2, we compare the minimum squared Euclidean distance between the transmission vectors $(d_{min})$, which is an important design parameter for quasi-static Rayleigh fading channels to optimize the error performance, of SIMO, SM, ESM, and QSM schemes. In all considered configurations, the average total transmitted energy is normalized to unity to make fair comparisons. It is interesting to note that ESM and QSM schemes achieve the same $d_{min}$ value for $4$ and $6$ bpcu transmissions. However, as seen from Fig. 2, QSM suffers a worse minimum Euclidean distance, as a result a worse error performance, compared to ESM scheme for higher spectral efficiency values, while the ESM scheme requires a more complicated transmitter with two RF chains. Finally, the results of Fig. 2 also prove that the relative $d_{min}$ advantage of IM schemes over classical SIMO scheme increases with increasing spectral efficiency, that is, IM techniques become more preferable for higher spectral efficiency values.

\begin{figure}[t]
	\begin{center}
		{\includegraphics[scale=0.7]{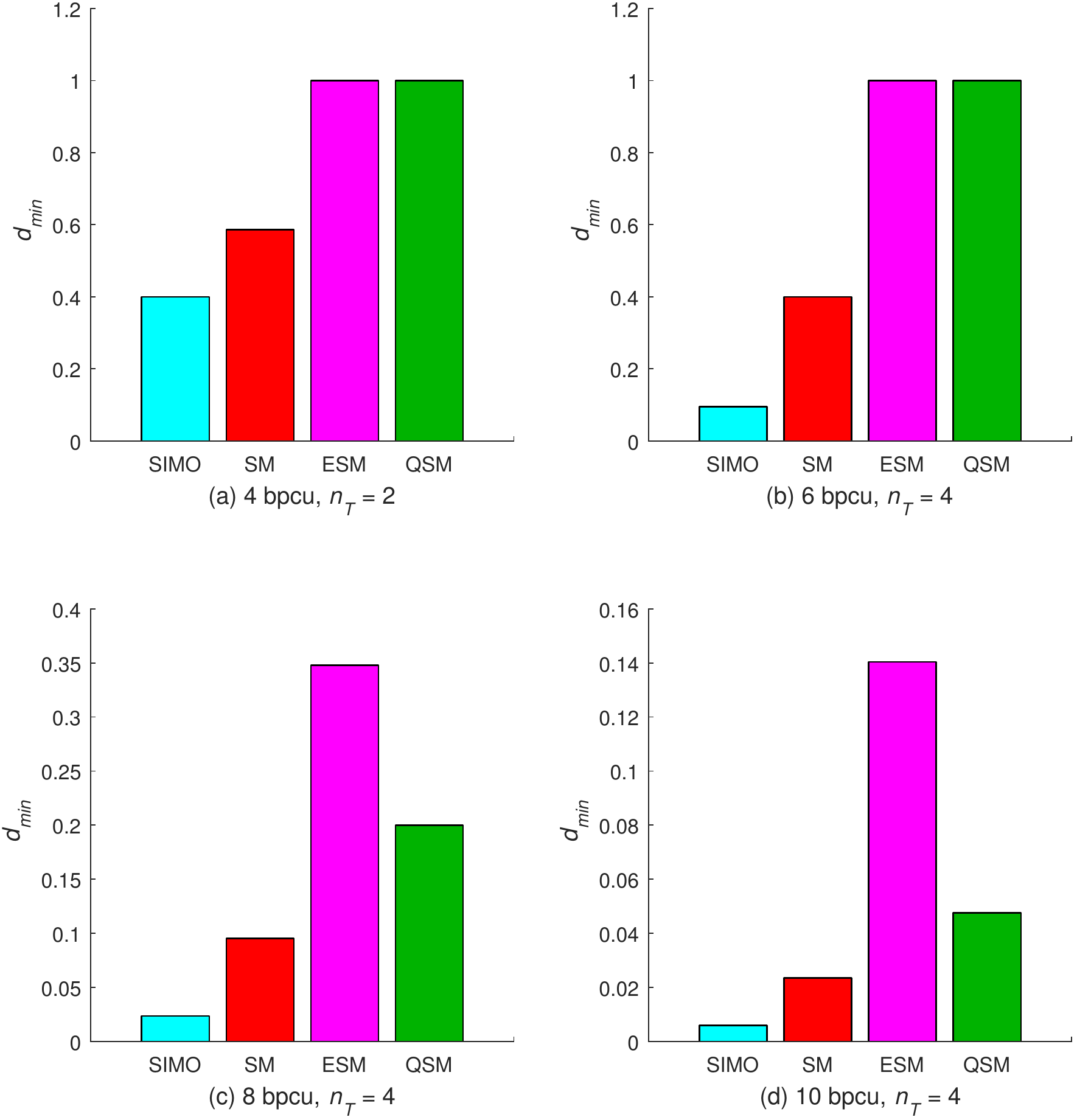}}
		\vspace*{-0.2cm}
		\caption{Minimum squared Euclidean distance $(d_{min})$ comparison of SIMO, SM, ESM and QSM schemes for different configurations (a) $4$ bpcu, $n_T=2$. SIMO:$16$-QAM, SM:$8$-PSK, ESM:QPSK/BPSK, QSM:QPSK. (b) $6$ bpcu, $n_T=4$. SIMO:$64$-QAM, SM:$16$-QAM, ESM:QPSK/BPSK, QSM:QPSK. (c) $8$ bpcu, $n_T=4$. SIMO:$256$-QAM, SM:$64$-QAM, ESM:$16$-QAM/QPSK, QSM:$16$-QAM. (d) $10$ bpcu, $n_T=4$. SIMO:$1024$-QAM, SM:$256$-QAM, ESM:$64$-QAM/$8$-QAM, QSM:$64$-QAM.  }
		\vspace*{-0.4cm}
	\end{center}
\end{figure}

\subsection{Massive Multi-user MIMO Systems with SM}
Massive MIMO concept, in which the BSs have tens to hundreds of antennas, is considered as one of the potential key technologies for 5G wireless networks due to its appealing advantages such as very high spectral and energy efficiency. While the initial studies on MIMO systems generally focus on point-to-point links where two users communicate with each other, practical MU-MIMO systems are gaining more and more attention to exploit the multiple antennas of a MIMO system to support multiple users simultaneously.

The extension of MIMO systems into massive scale provides unique opportunities for SM systems since it becomes possible to transmit higher number of information bits by the spatial domain with massive MIMO systems, even if the number of available RF chains is very limited. Although the spectral efficiency of SM systems cannot compete with that of traditional methods such as V-BLAST for massive MIMO systems, the use of IM concept for the transmit antennas of a massive MIMO system can provide an easy as well as cheap implementation solution thanks to the inherently available advantages of SM systems. Furthermore, SM is well-suited to unbalanced massive MIMO configurations in which the number of receive antennas are fewer than the number of transmit antennas \cite{Massive_SM}.

In Fig. 3(a), a massive MU-MIMO system is considered where $K$ users employ SM techniques for uplink transmission. Compared to user terminals with single antennas, additional information bits can be transmitted using SM without increasing the system complexity. GSM, ESM, and QSM techniques can be implemented at the users to further improve the spectral efficiency. At the BS, the optimal (ML) detector can be used at the expense of exponentially increasing decoding complexity (with respect to $K$) due to the inter-user interference. Low complexity near-optimal detection methods can be implemented as well by sacrificing the optimum error performance. Alternatively, SM techniques can also be used at the BS for downlink transmission as shown in Fig. 3(b). To support high number of users, the massive antennas of BS can be split into subgroups of fewer antennas where SM techniques can be employed for each user \cite{MU_SM}. For the specific case of two users, the data of User 1 can be mapped into antenna indices while the data of User 2 can be conveyed with $M$-ary signal constellations. GSM techniques can also be implemented at the BS to transmit the data of different users with either antenna indices and/or constellation symbols. 
 
\begin{figure}[t]
	\begin{center}
		{\includegraphics[scale=0.7]{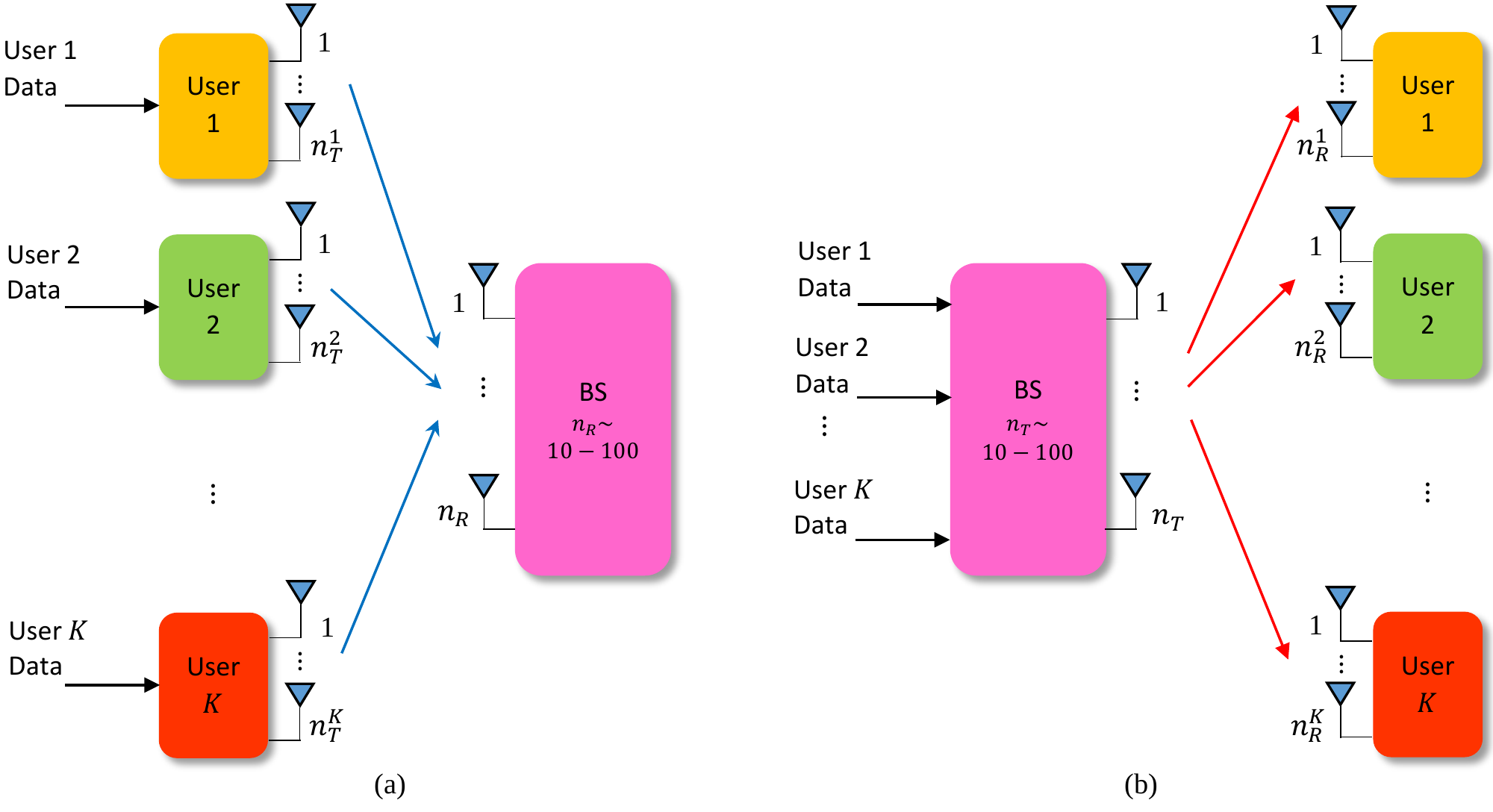}}
		\vspace*{-0.2cm}
		\caption{Massive MU-MIMO systems with SM (a) An uplink transmission scenario where User $k$ has $n_T^k$ transmit antennas available for SM and the BS has $n_R \sim 10-100$  receive antennas, (b) A downlink transmission scenario where User $k$ has $n_R^k$ receive antennas  and the BS has $n_T \sim 10-100$ transmit antennas available  for SM.   }
		\vspace*{-0.4cm}
	\end{center}
\end{figure}

\subsection{Cooperative SM Systems}

Cooperative communications, which allows the transmission of a user's data not only by its own antenna, but also by the active or passive nodes available in the network, has been one of the hot topics in the wireless communications field in the past decade.  Initially, cooperative communication systems have been proposed to create virtual MIMO systems for the mobile terminals due to the problems such as cost and hardware associated with the employment of multiple antennas in mobile terminals. However, due to the recent technological advances, multiple antennas can be employed at mobile terminals, and cooperative communications systems can efficiently provide additional diversity gains and high data rates by improving coverage. Consequently, relaying technologies have been incorporated into Long Term Evolution Advanced (LTE-A) standard for the purpose of increasing coverage, data rate, and cell-edge performance.

Considering the effective solutions provided by SM techniques and cooperative communications systems, the combination of these two technologies naturally arises as a potential candidate for future wireless networks. Due to the recent technological advances, cooperative SM systems can provide new implementation scenarios, additional diversity gains, and higher data rates without increasing the cost and complexity of the mobile and relay terminals. In the past few years, researchers have shown that SM techniques can be efficiently implemented for decode-and-forward (DF) and amplify-and-forward (AF) relaying based cooperative networks, distributed cooperation, and network coding systems. Readers are referred to \cite{SM_magazine_2} and \cite{DSM} and the references therein for further information on cooperative SM systems.

In Fig. 4, we consider four different cooperative SM system configurations where S, R, and D respectively stand for the source, relay, and destination node. In Fig. 4(a), a dual-hop network is considered, where SM techniques can be implemented at S and R with DF or AF based relaying techniques. The scenario of Fig. 4(a) is generally observed in practical networks where S and D cannot communicate directly due to distance or obstacles. In this relaying scenario, SM techniques can improve the energy and spectral efficiency of S compared to single antenna case, while multiple RF chains are required at R and D for signal reception. However, considering the uplink transmission from S to D, this would not be a major design problem. In Fig. 4(b),  a direct link from S to D exists and R can improve the quality of service of the transmission between S and D by employing different relaying methods.

\begin{figure}[!t]
	\begin{center}
		{\includegraphics[scale=0.75]{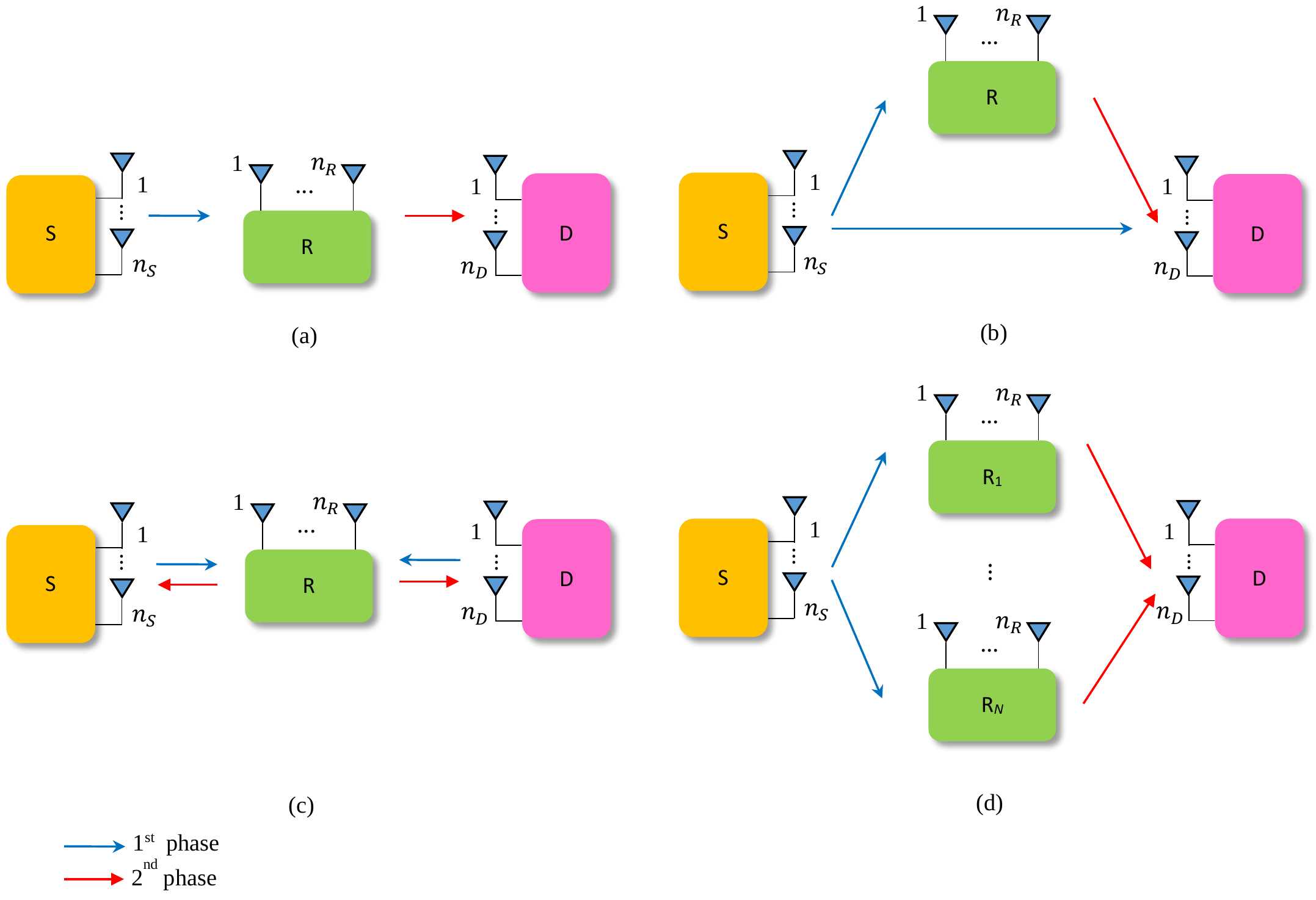}}
		\vspace*{-0.2cm}
		\caption{An overview of cooperative SM systems (a) Dual-hop SM (b) Cooperative SM (c) Network-coded SM (d) Multi-relay and distributed SM. $n_S,n_R$ and $n_D$ denote the number of antennas for source (S), relay (R) and destination (D) nodes, respectively.  }
		\vspace*{-0.4cm}
	\end{center}
\end{figure}

In Fig. 4(c), we take into account the two-way communications of S and D which is accomplished via R. Without network coding, the overall transmission between S and D requires four transmission phases (from S to R, R to D, D to R, and R to S) which considerably reduce the overall spectral efficiency. However, the two-way communications between S and D can be accomplished at two phases with network coding where in the first transmission phase, S and D  simultaneously transmit their signals to R using SM techniques. In the second transmission phase, R combines the signals received from S and D, then forwards this combined signal to S and D. The use of SM provides some opportunities for R such as transmitting one user's data with antenna indices and the other one's with constellation symbols. 

Finally, in Fig. 4(d), we consider a distributed cooperation scenario with $N$ relay nodes (R$_1,\ldots,$R$_N$). In the first transmission phase, S can use SM techniques to transfer its data to relays. In the second transmission phase, one or more relays cooperate and the indices of the activated relays can be considered as an additional way to convey information. This flexibility allows the relays to cooperate even if they have single antennas $(n_R=1)$. Furthermore, opportunistic relay selection is also an option for the network topology of Fig. 4(d), where the adaptively selected best relay takes part in transmission. For all cooperation scenarios described above, S and/or R can use GSM/ESM/QSM techniques to further improve the spectral efficiency as well as to obtain more flexibility in the system design.

\section{Index Modulation for OFDM Subcarriers: OFDM with Index Modulation}

IM concept can be considered for other communications systems apart from MIMO systems. For an instance, IM techniques can be efficiently implemented for the subcarriers of an OFDM system. OFDM-IM is a novel multi-carrier transmission scheme which has been proposed by inspiring from the IM concept of SM \cite{OFDM_IM}. Similar to SM, in the OFDM-IM scheme, the incoming bit stream is split into index selection and $M$-ary constellation bits. According to the index selection bits, only a subset of available subcarriers are selected as active, while the remaining inactive subcarriers are not used and set to zero. On the other hand, the active subcarriers are modulated according to the $M$-ary constellation bits. In other words, the information is conveyed not only by the data symbols as in classical OFDM, but also by the indices of the active subcarriers which are used for the transmission of the corresponding data symbols for the OFDM-IM scheme. 

Considering an OFDM system with $N_F$ subcarriers, one can directly select the indices of active subcarriers similar to IM technique used for the transmit antennas of an MA-SM system. However, the massive OFDM frames can provide more flexibility for the employment of IM techniques for OFDM-IM schemes compared to SM schemes. On the other hand, keeping in mind that $N_F$ can take very large values, such as $512,1024$ or $2048$ as in LTE-A standard, there could be trillions of (actually more than a googol $(10^{100})$ in mathematical terms) possible combinations for active subcarriers if index selection is applied directly. As an example, assume that we want to select the indices of $256$ active subcarriers out of $N_F=512$ available subcarriers, then, there could be $472.55 \times 10^{150}$ possible different combinations of active subcarriers, which turn the selection of active subcarriers into an almost impossible task. Therefore, for the implementation of OFDM-IM, the single and massive OFDM-IM block should be divided into $G$ smaller and manageable  OFDM-IM subblocks each containing $N$ subcarriers to perform IM, where $N_F=G\times N$. For each subblock, $K$ out of $N$ available subcarriers can be selected as active according to the  $p_1=  \lfloor \log_2 \binom{N}{K} \rfloor $ index selection bits where typical $N$ values could be $2,4,8,16,$ and $32$ with $1\le K < N$. Please note that classical OFDM becomes a special case of OFDM-IM with $K=N$, that is, when all subcarriers are activated. 

The block diagrams of OFDM-IM scheme's transmitter and receiver structures are illustrated in Figs. 5(a) and 5(b), respectively. As seen from Fig. 5(a), for each OFDM-IM frame, a total of
\begin{equation}
	m = pG= \left( \Bigl\lfloor \log_2 \binom{N}{K} \Bigr\rfloor + K \log_2 M \right)G 
	\label{OFDM}
\end{equation}
 bits can be transmitted where $p=p_1+p_2$ and $p_2=K \log_2 M$. In Fig. 5(a), $\mathbf{j}_g$ and $\mathbf{s}_g$ denote the vector of selected indices and $M$-ary data symbols with dimensions $K\times 1$, respectively. First, OFDM-IM subblock creator forms the $N\times 1$ OFDM-IM subblocks $\mathbf{x}_g,g=1,\ldots,G$, then the OFDM-IM block creator obtains the $N_F \times 1$ main OFDM-IM frame $\mathbf{x}$ by concatenating these $G$ OFDM-IM subblocks. After this point, $G\times N$ block interleaving can be performed to ensure that the subcarriers of a subblock are affected by uncorrelated wireless fading channels. Finally, classical OFDM procedures such as inverse fast Fourier transform (IFFT), cyclic prefix (CP) insertion, and digital-to-analog (DAC) conversion are applied.

Two different  index selection procedures are available for OFDM-IM: reference look-up tables for smaller $N$ values and combinatorial number theory method for higher $N$ values, where examples of these two methods are provided in Fig. 5(c). Similar to SM, the receiver of OFDM-IM has to determine the active subcarriers and the corresponding data symbols in accordance with the index selection procedure used at the transmitter. After applying inverse operations, first, the received signals are separated since the detection of different subblocks can be carried out independently. The optimum but high-complexity ML detector makes a joint search over possible subcarrier activation combinations and data symbols, while the low-complexity log-likelihood ratio (LLR) calculation based near-optimal detector determines the indices of the active subcarriers first, then, it detects the corresponding data symbols. The LLR detector calculates a probabilistic measure on the active status of a given subcarrier by considering the fact that the corresponding subcarrier can be either active (carrying an $M$-ary constellation symbol) or inactive. This detector is classified as near-optimal since it does not know the set of all possible subcarrier activation combinations.

\begin{figure}[!t]
	\begin{center}
		{\includegraphics[scale=0.75]{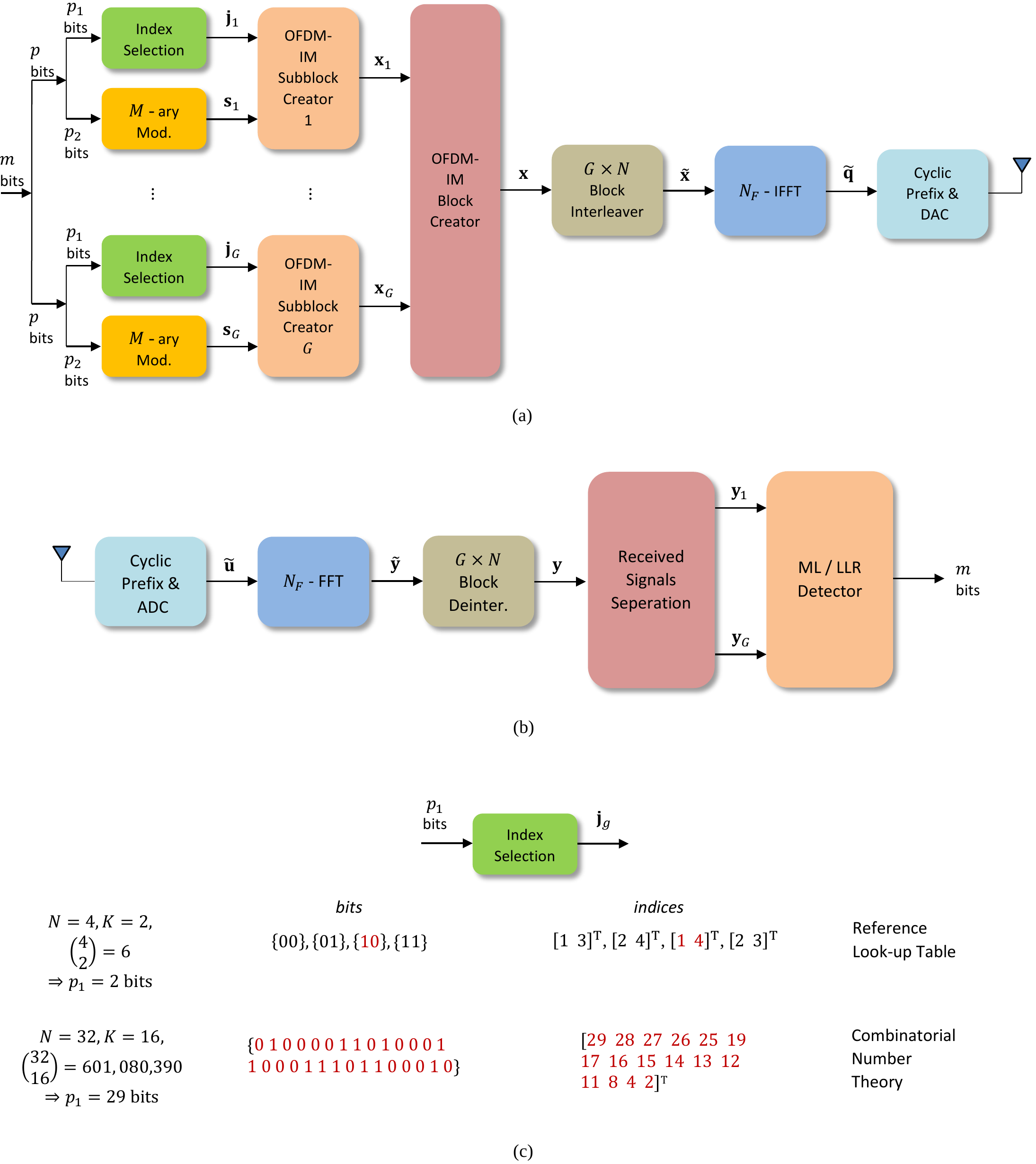}}
		\vspace*{-0.2cm}
		\caption{OFDM-IM system at a glance (a) Transmitter structure (b) Receiver structure (c) Two different index selection procedures.}
		\vspace*{-0.4cm}
	\end{center}
\end{figure}

It has been shown that OFDM-IM provides an interesting trade-off between error performance and spectral efficiency, and it offers some attractive advantages over classical OFDM. Unlike classical OFDM, the number of active subcarriers of an OFDM-IM scheme can be adjusted accordingly to reach the desired spectral efficiency and/or error performance. Furthermore, due to the information bits carried by IM, which have lower error probability compared to ordinary $M$-ary constellation bits, OFDM-IM can provide better bit error rate (BER) performance than classical OFDM for low-to-mid spectral efficiency values while it exhibits comparable decoding complexity using the near-optimal LLR detector. Furthermore, it has been recently proved that OFDM-IM also outperforms the classical OFDM in terms of ergodic achievable rate \cite{Rate_OFDM_IM}. 

Consequently, we conclude that due to its flexible system design with adjustable number of active subcarriers and its attractive advantages over OFDM, OFDM-IM can be a possible candidate not only for high-speed wireless communications systems but also for machine-to-machine (M2M) communications systems of 5G wireless networks which require low
power consumption.

\section{Recent Advances in OFDM-IM}
Subcarrier IM concept for OFDM has attracted significant attention from the researchers in recent times and it has been investigated in some up-to-date studies which deal with the error performance and capacity analysis, generalization, enhancement, and optimization of OFDM-IM, and its adaptation to different wireless environments. Interested readers are referred to \cite{MIMO_OFDM_IM} and the references therein for an overview of these studies. In this section, we focus on two recently proposed and promising forms of OFDM-IM: generalized OFDM-IM and MIMO-OFDM-IM systems.

\subsection{Generalized OFDM-IM Schemes}
Two generalized OFDM-IM structures (OFDM-GIM-I and OFDM-GIM-II) have been recently proposed  by modifying the original OFDM-IM scheme \cite{OFDM_GIM}. In OFDM-GIM-I scheme, the number of active subcarriers are no longer fixed and it is also determined according to the information bits. Considering the case of $N=4,K=2$ with binary PSK (BPSK) modulation $(M=2)$, according to (\ref{OFDM}), $ \lfloor \log_2 \binom{4}{2} \rfloor   + 2 \log_2(M) =4  $ bits can be transmitted per OFDM-IM subblock, that is, a total of $4\times 2^2=16$ subblock realizations can be obtained. On the other hand, considering all activation patterns $(K \in \left\lbrace 0,1,2,3,4\right\rbrace )$, which means that the number of active subcarriers can take values from zero (all subcarriers are inactive, $K=0$) to four (all subcarriers are active, $K=4$), as well as considering all possible values of $M$-ary data symbols, a total of $\sum\nolimits_{K=0}^{N} \binom{N}{K}M^K=81 $ possible subblock realizations can be obtained for which $\lfloor\log_2(81) \rfloor =6$ bits can be transmitted per OFDM-GIM-I subblock. As a result, the OFDM-GIM-I scheme can provide more flexibility for the selection of active subcarriers and can transmit more bits per subblock compared to OFDM-IM.

The OFDM-GIM-II scheme aims to further improve the spectral efficiency by applying IM independently for in-phase and quadrature components of the complex data symbols similar to the QSM scheme. In other words, a subcarrier can be active for one component, while it can be inactive simultaneously for the other component. Considering the case of $N=16,K=10$ with quadrature PSK (QPSK) modulation $(M=4)$, according to (\ref{OFDM}), $ \lfloor \log_2 \binom{16}{10} \rfloor   + 10 \log_2(M) =32  $ bits can be transmitted per OFDM-IM subblock. On the other hand, the OFDM-GIM-II scheme allows the transmission of $ \lfloor \log_2 \big(  \binom{16}{10}(\sqrt{M})^K \times \binom{16}{10}(\sqrt{M})^K  \big) \rfloor   =44  $ bits per subblock, which is $37.5$\% higher than that of OFDM-IM. Please note that the in-phase and quadrature components of a complex $M$-QAM symbol are the elements of a $\sqrt{M}$-ary pulse amplitude modulation (PAM) constellation, where a total of $ \binom{N}{K}\times (\sqrt{M})^K  $ realizations are possible per each component.

\subsection{From SISO-OFDM-IM to MIMO-OFDM-IM}
The first studies on OFDM-IM generally focused on point-to-point SISO systems, which can be unsuitable for some applications due to their limited spectral efficiency. More recently, MIMO transmission and OFDM-IM principles are combined to further boost the spectral and energy efficiency of the OFDM-IM scheme  \cite{MIMO_OFDM_IM}. Specifically, the transmitter of the MIMO-OFDM-IM scheme consists of parallel concatenated SISO-OFDM-IM transmitters (Fig. 5(a)) to operate over $n_T\times n_R$ MIMO frequency selective fading channels. At the receiver of the MIMO-OFDM-IM scheme, the simultaneously transmitted OFDM-IM frames are
separated and demodulated using a low-complexity MMSE detection and LLR calculation based detector which considers the statistics of the MMSE filtered received signals. It has been shown via extensive computer simulations that due to its improved error performance and flexible system design, MIMO-OFDM-IM can be a strong alternative to classical MIMO-OFDM, which has been included in many current wireless standards. 

\begin{figure}[!t]
	\begin{center}
		{\includegraphics[scale=0.8]{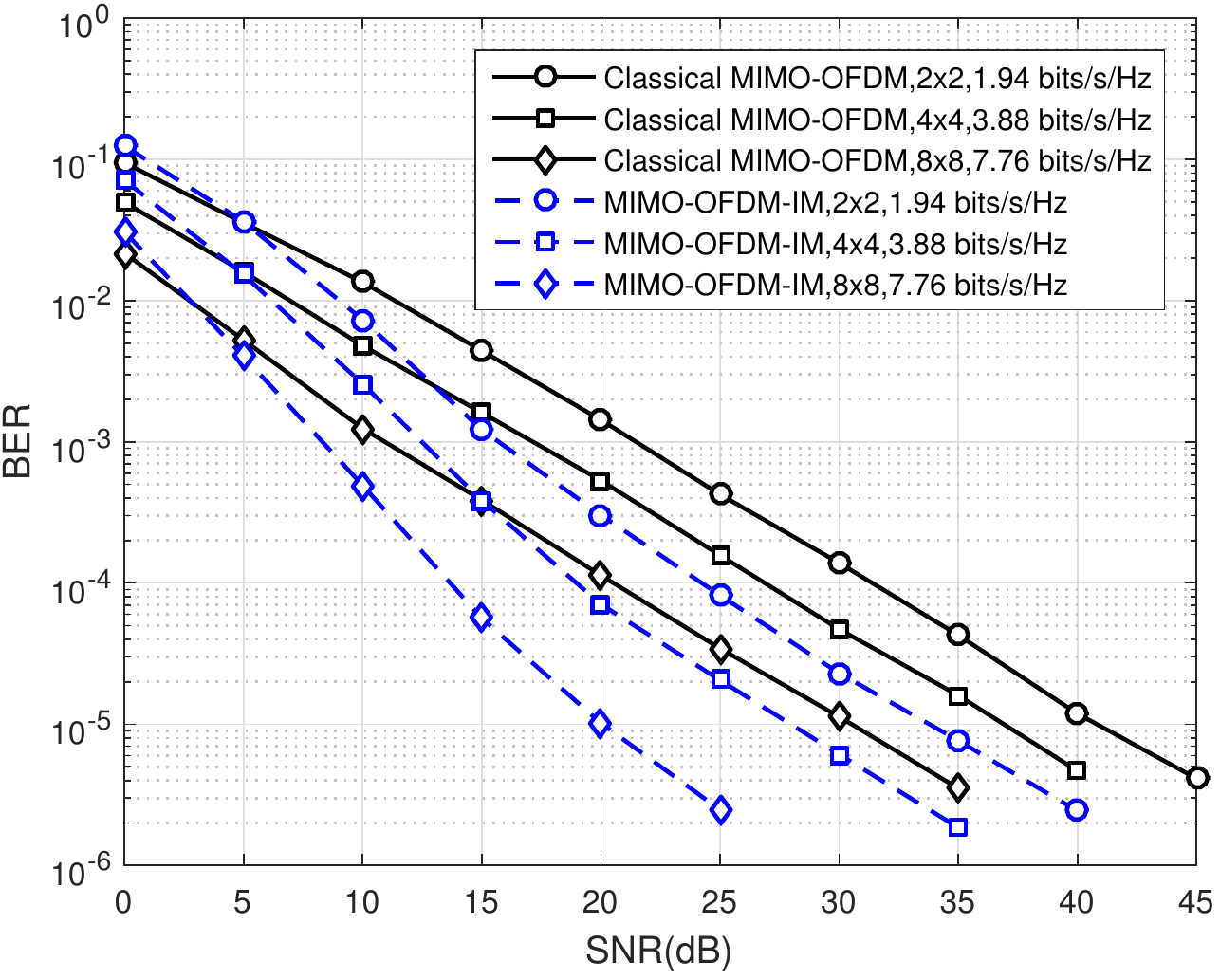}}
		\vspace*{-0.2cm}
		\caption{Uncoded BER performance of MIMO-OFDM-IM and classical MIMO-OFDM schemes for three $n_T \times n_R $ MIMO configurations: $2\times 2, 4\times 4$, and $8 \times 8$. OFDM system parameters: $M=2$ (BPSK), $N=4,K=2$, $N_F=512$, CP length $= 16 $, frequency-selective Rayleigh fading channel with $ 10 $ taps, uniform power delay profile, successive MMSE detection. The $3$\% reduce in spectral efficiency compared to single-carrier case $(n_T \log_2 M)$ is due to CP (Reproduced from \cite{MIMO_OFDM_IM} with permission).}
		\vspace*{-0.4cm}
	\end{center}
\end{figure}

In Fig. 6, the uncoded BER performance curves of the MIMO-OFDM-IM and classical V-BLAST type MIMO-OFDM schemes are given for three MIMO configurations where the same spectral efficiency values are obtained for both schemes. As observed from Fig. 6, significant signal-to-noise ratio (SNR) improvements can be obtained by the MIMO-OFDM-IM scheme compared to classical MIMO-OFDM to reach a target BER value. On the other hand, the generalization of MIMO-OFDM-IM for massive MU-MIMO systems remains an interesting research problem towards 5G wireless networks. 

Another recently proposed IM scheme, which is called generalized space-frequency index modulation (GSFIM)\cite{GSFIM}, combines OFDM-IM concept with GSM principle by exploiting both spatial and frequency (subcarrier) domains for IM. It has been shown that GSFIM scheme can also provide improvements over MIMO-OFDM in terms of achievable data rate and BER performance with ML detection for BPSK and QPSK constellations. However, the design of low complexity detector types is an open research problem for the GSFIM scheme.

\section{Conclusions and Future Work}
IM is an up and coming concept for spectral and energy-efficient next generation wireless communications systems to be employed in 5G wireless networks. IM techniques can offer low-complexity as well as spectral and energy-efficient solutions towards the single/multi-carrier, massive MU-MIMO, and cooperative communications systems to be employed in 5G wireless networks. In this article, we have reviewed the basic principles, advantages, recent advances, and application areas of SM and OFDM-IM systems, which are two popular applications of the IM concept. In Table II, the pros and cons of the reviewed IM schemes in terms of spectral efficiency, ML detection complexity, and error performance are provided. We conclude from Table II that IM schemes can be considered as possible candidates for 5G wireless networks due to the interesting trade-offs they offer among error performance, complexity, and spectral efficiency, while there are still interesting as well as challenging research problems need to be solved in order to further improve the efficiency of IM schemes. These research challenges can be summarized as follows:
\begin{itemize}
	\item The design of novel generalized/enhanced IM schemes with higher spectral and/or energy efficiency, lower transceiver complexity, and better error performance
	\item  The integration of IM techniques (such as SM, GSM, ESM, QSM, and OFDM-IM) into massive MU-MIMO systems to be employed in 5G wireless networks and the design of novel uplink/downlink transmission protocols
	\item The adaption of IM techniques to cooperative communications systems (such as dual/multi-hop, network-coded, multi-relay, and distributive networks)
	\item The investigation of the potential of IM techniques via practical implementation scenarios. 
\end{itemize}

 \setlength\extrarowheight{1.5pt}
\begin{table}[!t]
	\centering
	\caption{Pros and Cons of Several Index Modulation Schemes}
	\label{my-label}
	\begin{tabular}{l|p{3cm}||l|l|l|}
		\cline{2-5}
		& Scheme    & \begin{tabular}[c]{@{}l@{}} Spectral \\ efficiency\end{tabular} & \begin{tabular}[c]{@{}l@{}}ML detection \\ complexity\end{tabular} & \begin{tabular}[c]{@{}l@{}}Error\\ performance\end{tabular} \\ \hline \hline
		\multicolumn{1}{|l|}{\multirow{7}{*}{\begin{tabular}[c]{@{}l@{}}Single-\\ carrier\\ commun.\\ systems\end{tabular}}} & SIMO & Low & Low                                                                & Low                                                         \\ \cline{2-5} 
		
		\multicolumn{1}{|l|}{}                                                                                             & SM           & Moderate                                                       & Low$^*$                                                                & Moderate                                                    \\ \cline{2-5} 
		\multicolumn{1}{|l|}{}                                                                                             & GSM          & Moderate                                                       & Moderate$^*$                                                           & Moderate                                                    \\ \cline{2-5} 
		\multicolumn{1}{|l|}{}                                                                                             & MA-SM        & High                                                           & Moderate$^*$                                                           & Moderate                                                        \\ \cline{2-5} 
		\multicolumn{1}{|l|}{}                                                                                             & ESM          & High                                                           & Low                                                                & High                                                        \\ \cline{2-5} 
		\multicolumn{1}{|l|}{}                                                                                             & QSM          & High                                                           & Low                                                                & High                                                        \\ \cline{2-5} 
		\multicolumn{1}{|l|}{}                                                                                             & V-BLAST      & High                                                           & High$^*$                                                            & Moderate                                                    \\ \hline \hline
		
		\multicolumn{1}{|l|}{\multirow{7}{*}{\begin{tabular}[c]{@{}l@{}}Multi-\\ carrier\\ commun.\\ systems\end{tabular}}}  & OFDM         & Low                                                       & Low                                                                & Low                                                         \\ \cline{2-5} 
		\multicolumn{1}{|l|}{}                                                                                             & OFDM-IM      & Low                                                       & Moderate$^*$                                                                & Moderate                                                    \\ \cline{2-5} 
		\multicolumn{1}{|l|}{}                                                                                             & OFDM-GIM-I   & Moderate                                                           & High$^*$                                                           & Moderate                                                    \\ \cline{2-5} 
		\multicolumn{1}{|l|}{}                                                                                             & OFDM-GIM-II  & Moderate                                                           & High$^*$                                                           & Moderate                                                    \\ \cline{2-5} 
		
		\multicolumn{1}{|l|}{}                                                                                             & MIMO-OFDM-IM & High                                                           & High$^*$                                                               & High                                                        \\ \cline{2-5} 
		\multicolumn{1}{|l|}{}                                                                                             & GSFIM        & High                                                           & High                                                               & Moderate                                                    \\  \cline{2-5}
		
		\multicolumn{1}{|l|}{}                                                                                             & V-BLAST-OFDM    & High                                                           & Moderate$^*$                                                               & Moderate                                                    \\ \hline \hline

		\multicolumn{5}{l}{$^*$Lower complexity near/sub-optimal detection is also possible. }  \\[12ex]
	\end{tabular}
		\vspace*{-1.9cm}
\end{table}

 \newpage

\bibliographystyle{IEEEtran}
\bibliography{IEEEabrv,kitap_2016}

\vspace*{2cm}
\section*{Biography}
Ertugrul Basar [S'09, M'13, SM'16] received his B.S. degree with high honors from Istanbul University, Istanbul, Turkey, in 2007, and his M.S. and Ph.D. degrees from Istanbul Technical University, Istanbul, Turkey, in 2009 and 2013, respectively. He spent the academic year 2011-2012 at the Department of Electrical Engineering, Princeton University, New Jersey, USA. Currently, he is an assistant professor at Istanbul Technical University, Electronics and Communication Engineering Department and a member of Wireless Communication Research Group. He was the recipient of Istanbul Technical University Best Ph.D. Thesis Award in 2014 and won two Best Paper Awards. He is a regular Reviewer for various IEEE journals and served as a TPC Member for several conferences. His primary research interests include MIMO systems, index modulation, cooperative communications, OFDM, and visible light communications.


\end{document}